\documentstyle[referee]{l-aa}

\def\R#1{{\mathrm{#1}}}		
\def\Eq#1{{Eq.~(\ref{e:#1})}}	

\def\M#1{{\mathbf{#1}}}	
\def\T#1{{{#1}^{\bot}}}		
\def\d#1{{\R{d}{#1}}}		
\def\mdot{\!\cdot\!}		

\begin{document}

   \thesaurus{04  
      	{10.11.1; 
	 10.06.2; 
	 10.08.1; 
	 10.19.1; 
	 10.19.3; 
	 12.04.1}} 

\title{The distribution  of  nearby  stars in phase  space   mapped by
Hipparcos\thanks{Based   on    data   from  the  Hipparcos  astrometry
satellite}}

\subtitle{I. The potential well and local dynamical mass}

\author{M.~Cr\'ez\'e\inst{1,2}, E. Chereul\inst{1}, O. Bienaym\'e\inst{1} \and 
	C.Pichon\inst{3} } 

\offprints{M.~Cr\'ez\'e}

\institute{Centre de Donn\'ees astronomique de Strasbourg, CNRS URA1280 
	\\11 rue de l'Universit\'e, F-67000 Strasbourg, France
\and	IUP de Vannes,\\ 8 rue Montaigne, BP 561, 56017 Vannes Cedex, France
\and	 Astronomisches  Institut Univ. Basel,\\ Venusstrasse 7, 
	CH-4102 Binningen, Switzerland }

\date{Received / Accepted }

\maketitle

\begin{abstract}
          Hipparcos data provide   the   first, volume   limited   and
absolute  magnitude limited homogeneous  tracer of stellar density and
velocity  distributions  in the  solar  neighbourhood. The  density of
A-type stars more luminous than  $M_v=2. 5$  can be accurately  mapped
within a sphere of  125 pc radius,  while proper motions in  galactic
latitude provide the vertical velocity  distribution near the galactic
plane.  The potential   well   across the   galactic  plane is  traced
practically hypothesis-free   and  model-free.   The  local  dynamical
density comes out as  $\rho_{0}=0.076 \pm0.015~M_{\sun}~{\rm pc}^{-3}$ a
value well below all previous  determinations leaving no room for  any
disk shaped component of dark matter.


\end{abstract}

%

\section{Introduction}

All the data used here were collected by the Hipparcos satellite
(\cite{Hip97}). Individual stellar distances within  more than 125  pc
were obtained with an accuracy  better than 10$\%$ for almost all
stars  brighter  than $m_v  =  8.  $,   together with accurate  proper
motions.   Based on  these data, a  unique  opportunity is  offered to
revisit stellar kinematics and dynamics; any subsample of sufficiently
luminous stars is completely included within well defined distance and
luminosity limits, providing  a   tracer of the local   density-motion
equilibrium in the galaxy potential: a snapshot of the phase space. We
have selected  a series of  A-F dwarf samples  ranging from $M_v=-1.0$
down to $M_v=4.5$.  Completeness is  fixed within 50~pc over the whole
magnitude  range and within 125~pc   at  the luminous  end ($M_v  \leq
2.5$). In this series of  papers we shall  investigate such samples in
terms of density and velocity distribution small scale inhomogeneities
addressing the problem of cluster melting and phase mixing.

The  expected first order  departure  to homogeneity is the  potential
well across the galactic plane. This problem is well known in galactic
dynamics; it is usually referred  to as  ``the $K_z$ problem'',  where
$K_z$ means the force  law  perpendicular to  the galactic plane.  The
$K_z$ determination and    subsequent  derivation of  the   local mass
density $\rho_{0}$ has a   long history, nearly comprehensive  reviews
can be found in \cite{KerLB86} covering the subject before 1984 and in
\cite{Ku95}   since 1984.  Early  ideas were  given by Kapteyn (1922),
while Oort (1932) produced the first tentative determination.

The essence of this determination  is quite simple: the kinetic energy
of stellar motions in the z direction when stars cross the plane fixes
their capability  to  escape away from   the  potential well.  Given a
stellar population    at equilibrium in this  well,   its  density law
$\nu(z)$ and velocity distribution at plane crossing $f(w_0)$ are tied
to each  other via the $K_z$ or  the potential $\phi(z)$.  Under quite
general conditions  the relation that   connects both distributions is
strictly expressed by Eq. (\ref{eq1}).
\begin{equation}
		\nu(\phi) = 2 \int_{\sqrt{2\phi}}^\infty 
		{ f(|w_0|)~ w_0~{\rm d}w_0 \over \sqrt{w_0^2-2\phi} }
\label{eq1}
\end{equation}

This integral equation and its validity conditions are established and
discussed in detail by \cite{FuWi93} and \cite{FlyFu94}. There is no
specification as to the form of distribution  functions $f$ and $\phi$
except smoothness and separability of the z component.

Given $\nu(z)$ and $f(w_0)$,  $\phi(z)$ can be derived. Then according
to Poisson equation, the local dynamical density comes out as
\begin{equation}
	\rho_0 = {1 \over 4\pi G} ({\rm d}^2\phi / {\rm d} z^2)
\label{eq2}
\end{equation}

Early determinations of   $\rho_{0}$ (\cite{Oort60}) were  as  high as
$0.18\,M_{\sun}\,{\rm pc}^{-3}$ while  the total mass density of stars
and  interstellar  matter   would   not  exceed  $0.08\,M_{\sun}\,{\rm
pc}^{-3}$.  The    discrepancy  between  these   two   figures  latter
re-emphasised by  J.  Bahcall  and collaborators (1984ab,  1992), came
into the picture of   the dark matter  controversy   to make it   even
darker.   While the local  density of standard ``dark halos'' required
to maintain  flat rotation curves of galaxies   away from their center
would hardly exceed  0.01, such ten times  larger densities could only
be accounted  for in  rather   flat components which  composition  and
origin could  not be  the  same.  Following Bahcall's   claim, several
attempts were made to reduce the intrinsic inaccuracy of the dynamical
mass determinations  in particular \cite{BienRoCre87}, used a complete
modelisation of the galaxy evolution to tie the $K_z$ determination to
star counts,  \cite{KuGi89} used  distant  tracers of K type  stars to
determine not  the local  volume density  but the  integrated  surface
density  below  1   kpc,  \cite{FlyFu94} pointed  at  difficulties  in
relation with the  underlying dynamical approach.  Most concluded that
there was  no conclusive evidence for  a high dynamical density if all
causes of errors were taken into account.

The  local mass density  of observable  components (the observed mass)
can hardly  exceed $0.08 M_{\sun} {\rm  pc}^{-3}$ out  of which a half
stands  for stars which local  luminosity function  is reasonably well
known. The mass density of interstellar gas  and dust is not so easily
accessed since the bulk of it, molecular hydrogen,  can only be traced
via CO and  the ratio CO/$\rm{H_2}$ is not  so well established.  Also
interstellar components    are basically clumpy and   distances poorly
known: it is  not so easy  to define a proper  volume within which  a
density makes    sense. A conservative   range  for the  observed mass
density should be $ [0.06,0.10]$ including 0.04 for stars.

Nevertheless,   so  far the  discrepancy used   to   be charged to the
dynamical density since the   determination of this  quantity involved
many difficulties, most poorly controlled~:

\begin{itemize}
 
	\item Tracer homogeneity:  first of  all, the  determination
require  that a  suitable tracer can   be duly identified. A criterion
independent of velocity and distance should allow to detect and select
all stars matching the tracer definition. This  was hardly achieved by
spectral surveys  which    completeness  and selectivity degrade    as
magnitude grows.

	\item Stationarity: in order for the tracer to make dynamical
sense, it should  be in equilibrium  in the potential. Youngest  stars
which velocities reflect the kinematics of the gas at their birth time
do not fulfill this requirement.

	\item Undersampling:  one searches for  the bending of density
laws caused  by gravitation  along the z  direction. This  imposes not
only that tracer densities  can be determined, but  also that they can
be  determined   at scales  well below   the  typical length   of  the
bending.  This  means that  the tracer should  be  dense enough in the
region studied. Most determinations so far rely upon sample surveys in
the  galactic polar caps,  resulting in  ``pencil beam'' samples quite
inappropriate to trace the density  near the plane (\cite{Cre89}). When
the  bending is   traced at large   distances   from the   plane as in
\cite{KuGi89},   one gets information on the   surface density below a
certain  height. A comparison of this   surface density with the local
volume density involves modeling hypotheses. Another associated effect
can  be generated   by  a  clumpy  distribution   of the   tracer even
stationary if the tracer includes clusters or clumps the statistics of
densities and  velocities may be locally  biased. Other problems arise
from the velocity distribution sampling.

	\item  Systematic errors: out  of necessity before Hipparcos,
all previous studies used  photometric distances. Uncertainties in the
calibration  of absolute magnitudes resulted  in  systematic errors in
density determinations. This is particularly true for red giants which
are not at all a physically homogeneous family and show all but normal
absolute  magnitude distributions: it is striking that calibrations
based on different samples provide discrepant calibrations outside the
range  permitted by formal errors.  This produced  both biased density
distributions  and uncontrolled  effects  in  selecting volume limited
samples.

	\item  Random  errors:   distances  and  velocities  were also
affected by   random measurement  errors.  Even perfectly  calibrated,
standard errors  of  photometric distances based  on intermediate band
photometry would hardly get below 20\%.

\end{itemize}

Hipparcos data solve nearly all  the problems quoted above: instead of
``pencil beam'' tracers at  high galactic latitudes which distribution
is only indirectly related to the local  $K_z$, what we  get here is a
dense probe inside the potential well.

The samples   were  preselected  within magnitude  limits   inside the
completeness  limit       of     the   Hipparcos       survey  program
(\cite{TuCri86}). It includes all stars brighter than $m_v$ = 8.0 (7.9
at low latitudes)  with spectral types earlier  than G0, stars with no
or poorly defined spectral types were included provisionally.

Within this coarse preselection, sample stars were eventually selected
on the basis   of their  Hipparcos  derived   distance, magnitude  and
colour. So the only physical criterion is absolute magnitude while the
sample  is distance limited   on  the basis of individual  parallaxes.
Within 125  parsecs, Hipparcos parallaxes provide individual distances
with accuracies better than 10 percent for  over ninety percent of the
sample.  Distances are  individual  and free from  calibration errors.
Samples are dense, typical inter-star   distances are less than  10~pc.
 This is suitable to trace density  variations which typical scales are
of the  order of hundred parsecs. Thanks  to this  high tracer density,
difficulties    related  to  clumpiness  can    be  monitored  through
appropriate  analysis    of  local   density residuals.   Stationarity
considerations can also be monitored  since W velocities are available
for stars everywhere in the volume studied.

The main characteristics of the tracer samples are reviewed in section
2. In order to take  full benefit of this  new situation, an  original
method has  been developed to analyze  the  densities. The statistical
aspects  of this method based  on single star  volumes  (volume of the
sphere extending to the nearest neighbour)  are described in section 3
yielding a model free view of the potential  well. Then an estimate of
the  local  mass   density    of matter  is  produced   (section   4).
Consequences in  terms   of    galactic structure  and   dark   matter
distribution are reviewed in conclusion (section 5).

%

\section{The Hipparcos Sample}

\subsection {Sampling definition}

The sample was pre-selected from  the Hipparcos Input  Catalogue
(\cite{Hic92}),  among the   ``Survey stars''.  The  Hipparcos  Survey
delimitation is described in  full  detail in Turon \& Crifo (1986).   For
spectral types earlier  than G5    the limiting  magnitude is $    m_v
\leq~7.9+1.1sin\vert b\vert$. It  is  complete within those limits  so
far as   the apparent magnitudes   were  known.  Inside  the Hipparcos
Survey,  the phase space tracer sample  has  been first given a coarse
limitation:   spectral types  later    than A0 and   earlier  than G0,
luminosity   class V through IV.   Stars   with uncertain or undefined
luminosity classes were kept in the sample at this stage.  Within this
pre-selection,  the  final choice was   based  on Hipparcos magnitudes
($m_v  <  8.0$),  colours ($-0.1~\leq~B-V~\leq~0.6$)  and   parallaxes
($\pi~\leq~8~{\rm  mas}$).   Then a series of   subsamples were cut in
absolute magnitude  and distance in such a  way that each subsample is
complete  for the adopted colour  range within a  well defined sphere.
The  resulting sample characteristics are  summarized  in Tab.(1). The
sample designation system  is straightforward: h125 means stars within
125  pc, under the apparent  magnitude condition  $m_v < 8.0$, imposes
not to consider stars less luminous than $M_v \leq 2.5$.  All stars of
sample  h125 closer  than  100 pc are included  in  sample H100 and so
on. Samples  named k... come  from the corresponding h...  samples but
the identified cluster stars have been removed. The typical density of
each sample estimated as the inverse of the average single star volume
is  given in  the last  column  of Table \ref{tab1}. The corresponding
typical interstar distance of e.g. tracer h125 is 8.8~pc.

\begin{table}
\begin{center}
\caption {Main    sample characteristics.  M$_{v\   lim}$ the limiting
absolute magnitude, d$_{lim}$  the  limiting distance, N the  number of
stars, $<\nu>$ the mean number density.}
\label{tab1}
\small
\begin {tabular}{lcrccc}
\hline
\hline
 Sample & M$_{v\ lim}$ &d$_{lim}$ &N &$<\nu>$\\
 & & (pc)& &(10$^{-5}\cdot$pc$^{-3}$)\\
\hline
h125 & $\leq$ 2.5 & 125.9  & 2977 &  48\\ 

k125  & $\leq $ 2.5  &  125.9   & 2643  & 42\\

H100  & $\leq $ 3.0  &100.0 &  2677 & 88\\

K100  & $\leq $ 3.0 & 100.0  & 2385  &  77\\

H80   & $\leq $ 3.5  & 79.4  & 2336 &  152\\
\hline
\hline
\end{tabular}
\end{center}
\end{table}

\subsection { Completeness} 
With this  method,  the sample limits do  not  rely upon pre-Hipparcos
data except for the  faintest stars next to the  edge of each  sphere,
some of  those might have  been missed due to  the limited accuracy of
pre-Hipparcos magnitudes. The method may  look a naive one since there
are now sophisticated  algorithms to correct  estimations for censored
data, (see for instance \cite{Ratna91}) but it is the only one
that does  not   make   assumptions as  to  the    absolute  magnitude
distribution of the stars. All previous investigations did assume that
the absolute  magnitudes    of stars sharing  a    certain spectral or
photometric  characteristic are  normally  distributed around a common
mean value.

%
   \begin{figure}[htbp]
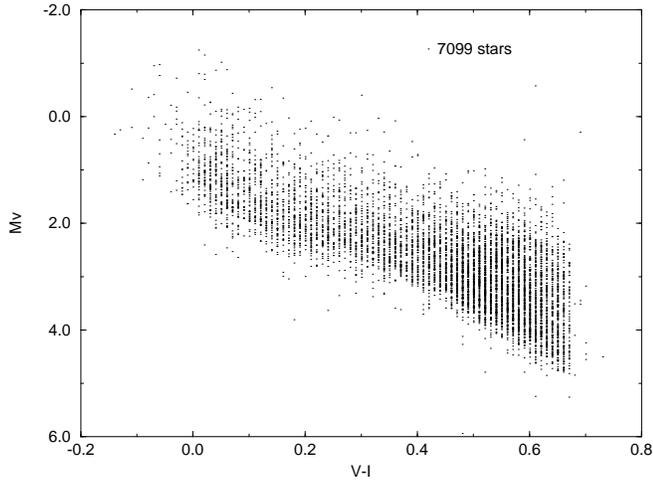

      \vspace{7cm}
	\special{ hscale=40 vscale=40 hoffset=-20 voffset=+220.
	hsize=450 vsize=450 angle=-90.  psfile=6440.f01 }
      \caption{ $M_v/V-I$  diagram  for the global  tracer sample (see
section 2.1).}
         \label{FigStaMVI}
   \end{figure}

%
   \begin{figure}[htbp]
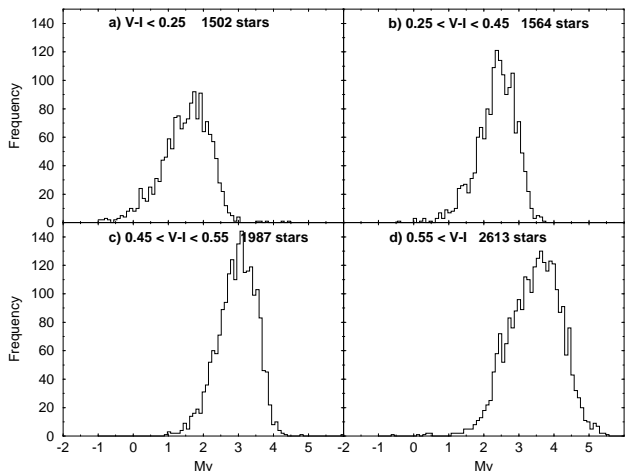

      \vspace{7cm}
	\special{ hscale=35 vscale=35 hoffset=-5 voffset=230.
	hsize=250 vsize=250 angle=-90.  psfile=6440.f02 }
      \caption{Absolute magnitude distribution  of  tracer stars in  a
series of colour range.} 
   \label{FigStaVI} \end{figure}

 Figures   \ref{FigStaMVI}  and   \ref{FigStaVI} give  a  quantitative
illustration   of   the   risks   involved:   the absolute   magnitude
distributions  of  dwarfs  within four  V-I   ranges show  an  obvious
asymmetry while in the reddest part, the luminosity distribution spans
over more than two magnitudes.  This means that however accurately the
apparent magnitude can be measured,  the uncensoring correction  would
be much more uncertain than other error factors.

\begin{figure}[htbp]
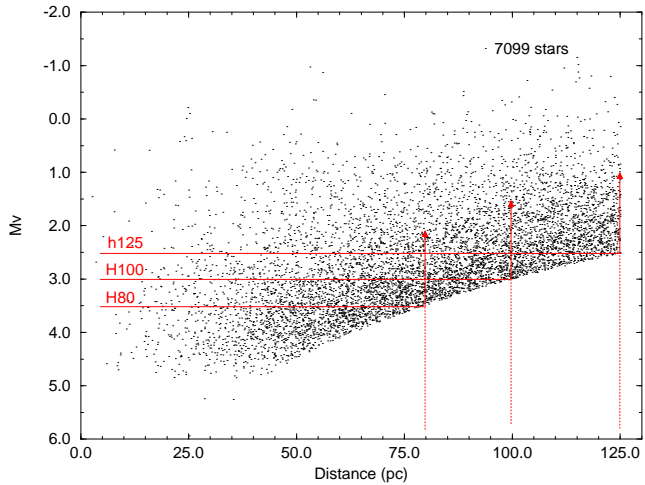

      \vspace{7cm}
	\special{ hscale=40 vscale=40 hoffset=-20 voffset=+220.
	hsize=450 vsize=450 angle=-90.  psfile=6440.f03 }
      \caption{  $M_v$/distance diagram  for the global  sample.  See
Table \ref{tab1} for sample definitions.}
         \label{FigStaMvD}
   \end{figure}

With  the approach  adopted   here,  Fig. \ref{FigStaMvD}  shows   the
$M_v$/distance distribution of the  global   sample, the neat   border
follows   simply the line $m_v=8$;  the  censoring effects are maximum
along  this line. Our  subsamples are cut  along lines parallel to the
axes of this plot, so they reach the censoring line only at the bottom
right corner of each zone.
%

Since the  accuracy  of   Hipparcos  magnitudes  is  far  beyond   the
necessities of this study,  the sampling biases  can only result  from
two effects:  the parallax errors  which, however  unprecedently small
are still of the order of $10\%$ beyond 100 pc, and  the stars lost at
the time  of the early  selection  due to  the inaccuracy  of apparent
magnitudes available then.  Concerning  the second effect, it can only
result in stars being omitted near the edge of the sample. The residual
effect   of apparent magnitude   incompleteness  can  only affect  the
faintest magnitudes   at the edge of  each  sample  sphere,  since the
single star volume used   in sections 3 and   4 as statistics  for the
density estimation  are local  statistics,   it is  easy to keep  such
effect under  control in the study of  residual, furthermore this kind
of incompleteness   if important would    bias the estimation  towards
making the potential well even shallower than observed.

\subsection {Distance errors}

\begin{figure}[htbp]
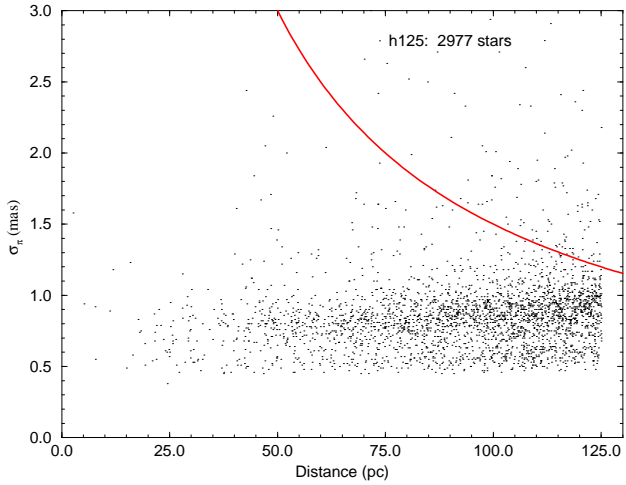

      \vspace{7cm}
	\special{ hscale=40 vscale=40 hoffset=-20 voffset=+220.
	hsize=450 vsize=450 angle=-90.  psfile=6440.f04 }
      \caption{  Formal  parallax  standard  errors  versus  distance.
	Distance  errors are larger   than  15$\%$ for  stars  above the
	line. }
         \label{FigStaErdis}
   \end{figure}

The  other  possible source  of  bias  left is generated   by distance
errors.  Figure \ref{FigStaErdis}  shows the distribution  of parallax
errors  versus distance for sample h125  which, extending to 125 pc is
the  most   severely affected by  distance errors.    This  is still a
moderate effect:  one can see  on the  figure that  only a handful of
parallax errors exceed  15$\%$. Parallaxe errors are Hipparcos formal
errors. All feasible  external tests  able  to check  the actual accuracy show
that   these formal  errors are  if   something  a  little pessimistic
(\cite{Arenou95}) with respect    to previous studies  of  the density
trends.

%

%
%

\section{Mapping the tracer density and the potential well }

The essential choice of  this investigation is  to stick as closely as
possible to observations. Whatever the velocity distribution $f(w_0)$,
once an homogeneous tracer has been  defined, that is once a selection
criterion  has been defined  uncensored with respect to velocities and
positions, Eq.  (\ref{eq1})  expresses the  correspondence between the
tracer density  law and the potential as  a plain change of variables.
So if  there is a  potential  well to  be  found in the data,  it must
appear as a density peak. If we can  map the density trend model free,
we'll get a picture of the potential well.

\subsection {Single star volume statistics}

The tracer should  be dense enough for  the typical  distances between
tracer stars to be small with respect to  the typical scale of density
variations. Under this assumption, at  a place were the tracer density
is $\nu$, the observed number of stars counted in any probe volume $V$
is a Poisson variate with expectation $ V \cdot \nu $. Defining $v$ as
the single star volume around  one specific star  (that is the  sphere
extending to the nearest neighbour),  and introducing the quantity $ x
= v \cdot \nu $ we get a new variate which expectation  is 1 and which
probability distribution is exponential  (Eq. (\ref{eq3})).  It is the
probability distribution  of the  distance   between two events  in  a
Poisson process under unit density.


\begin{equation}
{\rm d} P(x) = \exp(-x)~{\rm d} x 
\label{eq3}
\end{equation}

and the probability distribution of the single star volume $v$ is


\begin{equation}
{\rm d} P(v) = \nu \cdot \exp(- \nu \cdot v) ~{\rm d} v 
\label{eq4}
\end{equation} 

so the expectation of  the single star  volume is $1/\nu$, which makes
it a  suitable local statistics  for the  density. It  is not a  novel
approach to use  statistics  based on  nearest  neighbour distances to
investigate densities, but it is quite appropriate here since it turns
out to provide  a parameter free  Maximum Likelihood estimator for the
density by plain moving average of  single star volumes.  Also, single
star  volumes being  a  local measure of  the  inverse density at  the
position   of   each star, they   offer   the possibility to calculate
individual density residuals.  Checks can be made that deviations from
average or model predicted values are  randomly distributed and do not
show unmodeled systematic trends.

\subsection {Maximum Likelihood estimator}

In a constant density sample, the plain average of single star volumes
 is a maximum likelihood statistics for the  density: let $v_i$ be the
 single star volume around star   i, according to Eq. (\ref{eq4})  the
 log-likelihood of   a  sample of n  stars  $(i=1,n)$  under  constant
 density $\nu$ is simply given by

\begin{equation}
\log L = \sum_{i=1,n} ( \log (\nu) - \nu \cdot v_i )
\label{eq5}
\end{equation} 

and the Maximum Likelihood is reached for 
\begin{equation}
{\rm d} \log L/{\rm d}\,\nu = 0 \\
\label{eq6}
\end{equation}

which obvious solution is: 

\begin{equation}
n/\nu = \sum_{i=1,n} v_i \\
\label{eq7}
\end{equation}


\begin{equation}
\nu = 1 / <v_i>
\label{eq8} 
\end{equation}

this  makes the moving average  of $v_i$ along  a  parameter a maximum
likelihood  mapping of the   density variations along  this parameter,
provided only that variations along other parameters are negligible.

\subsection {Residual statistics}

  Furthermore, given a model or  an estimate of $\nu$  near star i the
quantities $x_i  = \nu \cdot v_i$  are local  ``residuals'' of the fit
and  their distribution  under   the assumed  model  should be exactly
given, parameter-free, by   Eq.  (\ref{eq3}). So  the distribution  of
residuals provides  an immediate  test of the  validity of   the model
including the fact that  the density is smooth  and do not include too
many clusters or voids. This distribution can be tested over the whole
sample or over any subset chosen to explore neglected parameters (e.g.
the completeness near the edge of the sampling volume, completeness in
apparent magnitude).     In  order  to visualize  the    agreement (or
disagreement)   of the observed     distributions  we have   used  two
representations: the histogram of the log residuals $y = log(x)$ which
distribution according to Eq. (\ref{eq3}) must be given by
\begin{equation}
{\rm d} P(y) = \exp [y - \exp(y)]{\rm d}y \\
\label{eq9}
\end{equation}

and the cumulative distribution of x: 


\begin{equation}
	H(X) = \int_0^X  \exp(-x)\,{\rm d}x = 1 -\exp(-X) \\
\label{eq10}
\end{equation}

Let $n_i$ be the rank of residual $x_i$ in a sample  of N values, then
according to Eq. (\ref{eq10}),


\begin{equation}
	n_i/N = 1-\exp(-x_i)\\ 
\label{eq11}
\end{equation}
	

\begin{equation}
	- \log(1- n_i/N) = x_i \\
\label{eq12}
\end{equation}

In the following the residuals  have been plotted as $-  log(1-n_i/N)$
against  $x_i$ producing what  statisticians  use to  call exponential
probability plots. On such  plot, observed residual distributions that
fail   matching the expected  exponential    at any scale sign   their
existence by departures from the diagonal.

\subsection {Edge effect}

\begin{figure}[htbp]
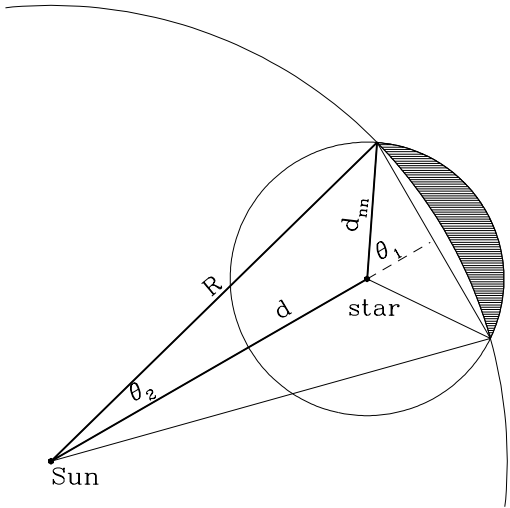

\vspace{8cm}
\special{ hscale=50 vscale=50 hoffset=-20 voffset=-80.
hsize=300 vsize=300 angle=0.  psfile=6440.f05 }
\caption{ Edge effect: geometry of the volume correction (shaded area) 
due to the part of the nearest neighbour volume outside
the completeness volume.}  \label{FigEdgeEffect} \end{figure}

Around stars close to the edge of  the completeness sphere, the single
star volume is   not anymore the full    volume of the  sphere to  the
nearest  neighbour.  The volume must be  corrected for the part of the
sphere outside  the  completeness  volume (Fig.  \ref{FigEdgeEffect}).
Defining  $d_{nn}$  the distance  to the  nearest   neighbour, $d$ the
Sun-star distance    and $R$ the radius  of    the sample  sphere, and
defining  the two  spheres, one centered  at sun  position with radius
$R$, the other  at star position  with radius $d_{nn}$,  the corrected
volume is calculated as $V_{cor}$ according to the following formulae.
Angles $\theta_1$  and $\theta_2$ are the  half apertures of the cones
intercepting the  intersection of   the two  spheres  (summit at   sun
position for $\theta_2$, at star position for $\theta_1$).

\begin{equation}
        \cos \theta_2 = ( d^2 + R^2 - d_{nn}^2 ) / (2 d R)
\label{eq13}
\end{equation}
\begin{equation}
        \sin \theta_1 = ( R/d_{nn} ) \sin \theta_2
\label{eq14}
\end{equation}
\begin{equation}
        V_1 = (\pi/3) d_{nn}^3 ( 2 - 3 \cos\theta_1 + \cos^3 \theta_1 )
\label{eq15}
\end{equation}
\begin{equation}
        V_2 = (\pi/3) R^3 ( 2 - 3 \cos\theta_2 + \cos^3 \theta_2 )
\label{eq16}
\end{equation}
\begin{equation}
        V_{cor} = (4/3) \pi d_{nn}^3 - (V_1 - V_2)
\label{eq17}
\end{equation}

%

\subsection {Mapping the density}

\begin{figure}[htbp]
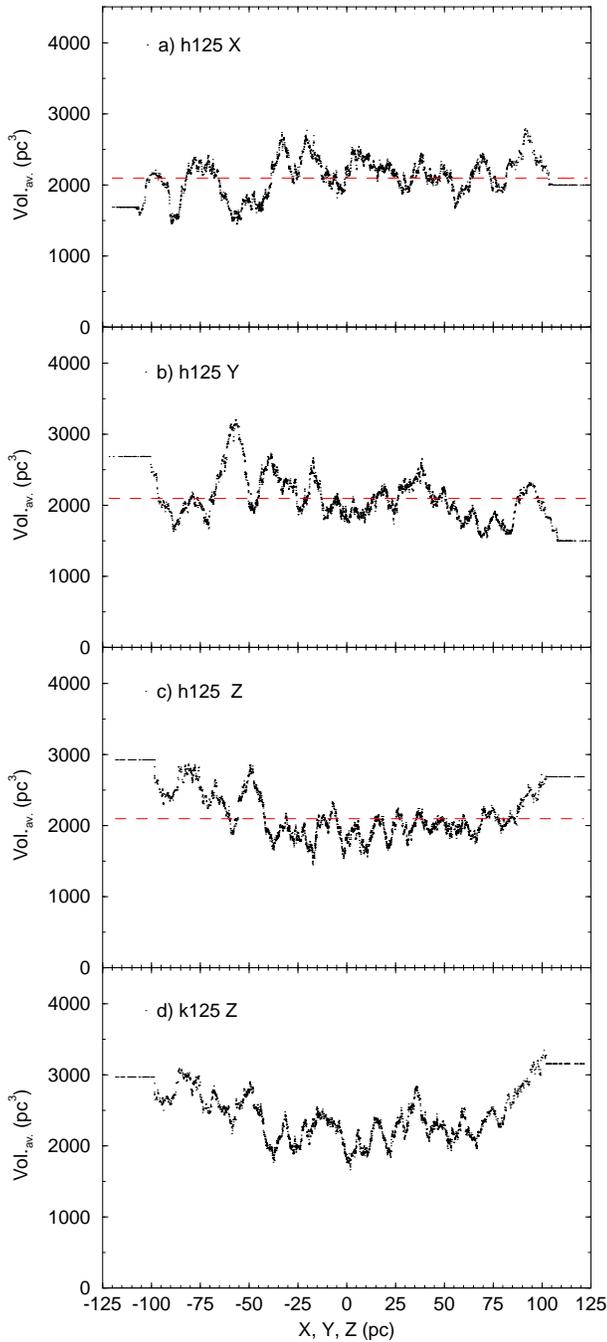

      \vspace{18cm}
	\special{ hscale=80 vscale=80 hoffset=-20 voffset=-50.
	hsize=450 vsize=550 angle=0.  psfile=6440.f06 }
      \caption{ Single star  volumes for  A stars  with $Mv \leq  2.5$
within 125  pc  (sample h125)   volumes in  cubic  parsecs are  moving
averaged along {\bf a)} the x ( l = 0, b = 0), {\bf b)} y  (l = 90°, b
= 0)  and {\bf  c)}   z (b =  90°)  directions.  Moving averages   are
performed over 101 stars, so that  random fluctuations are roughly one
order  of  magnitude below that   of  raw  data.  {\bf d)}  shows   in
comparison the same kind of plot for sample k125 axis z, where cluster
stars have  been removed  from sample   h125. The bending  due  to the
potential well in the galactic plane is clearly visible on plots c and
d. }
         \label{Momo125}
   \end{figure}

\begin{figure}[htbp]
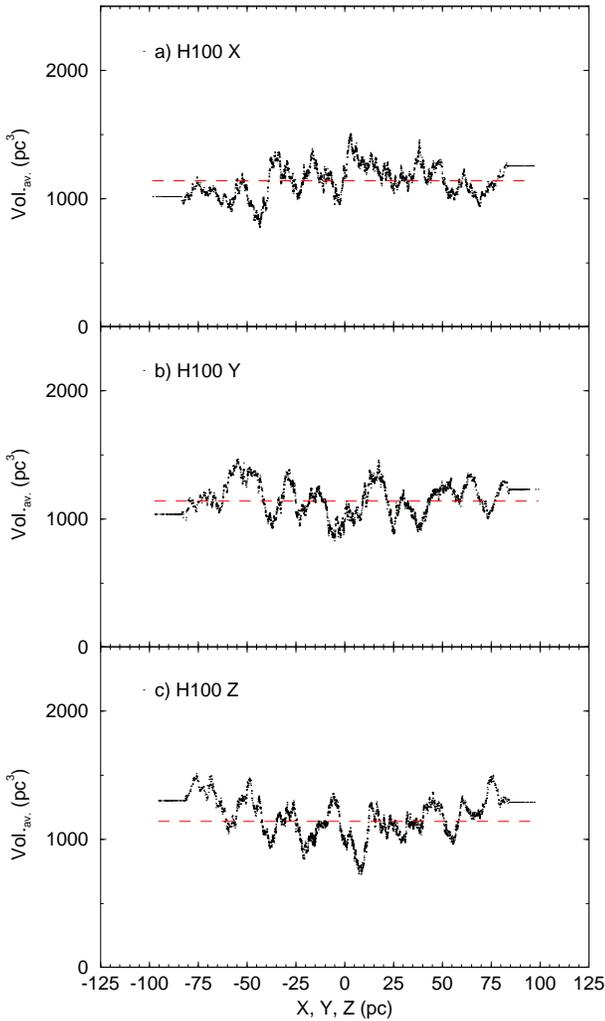

      \vspace{14cm}
	\special{ hscale=80 vscale=80 hoffset=-20 voffset=-50.
	hsize=450 vsize=450 angle=0.  psfile=6440.f07 }
      \caption{ same as Fig. \ref{Momo125} sample H100.  }
         \label{Momo100}
   \end{figure}

\begin{figure}[htbp]
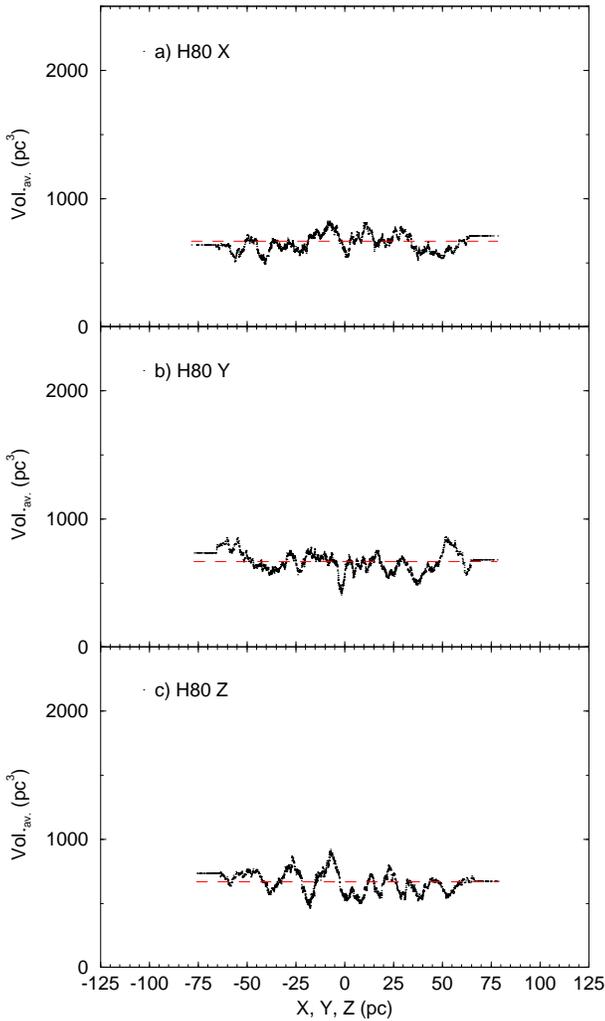

      \vspace{14cm}
	\special{ hscale=80 vscale=80 hoffset=-20 voffset=-50.
	hsize=450 vsize=450 angle=0.  psfile=6440.f08 }
      \caption{ same as Fig. \ref{Momo125} sample H80.  }
         \label{Momo80}
   \end{figure}

Single star volumes which are  unbiased local estimates of the inverse
tracer density have been computed for all the tracer samples described
in Table \ref{tab1}.  Defining x,y,z as a set of cartesian coordinates
centered at  the  sun  position,  with  z  positive  pointing  towards
$b=+90^\circ$, x positive towards the galactic  center ($l = 0^\circ$)
and y towards $l =  90^\circ$, each series of  values was sorted along
x,   y and z successively and   each resulting set  smoothed by moving
average  along the sort parameter.    Single star volumes are averaged
over 101  neighbouring   stars,  producing    parameter  free  maximum
likelihood   inverse  density profiles.  Figures \ref{Momo125} through
\ref{Momo80} show  the  resulting profiles along  x, y  and z  for our
three  most accurate and extended samples   (namely samples h125, H100
and H80, according to definitions of section 2.1).
In these figures the x and y plots are given only as references: since 
the sampling definition has spherical symmetry, any systematic effect along 
the z axis which would trace a sampling artifact rather than a real trend 
of the density would appear as well along the x and y axis.

%

%
%

%

%

%
%

At first glance only  do the h125 and H100  profiles along the  z axis
 show significant bending. Expectedly the youngest subsamples with the
 smallest   velocity    dispersions  are   also   the   most sensitive
 tracers. This is  a  direct mapping of the  potential  well. The next
 section will  be  dedicated to deriving  quantitative consequences of
 this observation. Beforehand the result  should be scrutinized in the
 light   of two  possible  criticisms: biases   in the inverse  volume
 estimates at  the  edges of the  sample  and clumpiness effects among
 poorly mixed A stars.

Edge  effects potentially related  to  completeness and/or to distance
errors can   be easily controlled: would  they  be responsible for the
main part of the observed  bending they would appear as  well on the x
and y plots. There is no such thing visible even at marginal level. In
addition, any significant incompleteness at the  edges would result in
a density decrease driving the single star volume statistics up at the
edges. As  a    result, the  dynamical  density  derived   below, once
corrected for this effect, would be even smaller.

A more serious problem comes from  the clumpiness of the distribution: 
young  stars that dominate the most  luminous samples are likely to be 
partly concentrated in  clusters or clumps  on scales of a few parsecs 
to a few ten parsecs.  Such clumps  must be suspected to  distort the 
density  profile. 
The fluctuations observed at small scales on figure 6 through 8 are not 
significant in this respect, they are the result of random fluctuations
of single star volumes smoothed by the moving average process. The moving
average introduces a strong self correlation in the series. As a result
random fluctuations fake systematic ones on scales which are connected with
the z ranges spanned by 101 neighbouring stars, but do not reflect these
ranges in any simple way (it has something to do with the probability
of getting one, two, three ... extreme peaks within adjacent strips).
Similar pictures are obtained on strictly uniform random simulations.

The only place where to check this effect is the statistics of residuals.
Real clumps which are not plain random fluctuations in a uniform density 
are expected to sign their presence not by their local density which
may not be very high locally on the average, but by the frequency of
corresponding single star volume residuals (see section 3.3 for the 
definition of these residuals. Such an overfrequency is
clearly visible on  the bottom left wing of the histogram on figure 9a.
In terms of cumulative distribution it is even more clear on figure 9b.
A separate analysis of clustering  has been performed by
means of wavelet analysis  (Chereul et al.  to  appear as paper II  of
this series) showing that no more than 8$\%$ out  of stars in the h125
sample  are involved in clumps.   Removing only cluster  members
identified  through proper motions we  get sample  k125. The z inverse
density profile  of  this  new sample  is  plotted  on figure 6d,  the
profile is not strongly modified by removing cluster stars.  On  the 
residual  plot (figure  9c)  the  residual anomaly  at small  scales 
vanishes definitely.  So the  inverse  density profiles turn  out quite 
robust to clumpiness effects. Furthermore we shall see in the following 
(section 4.2 and figure 13) that local mass density estimates based on the 
two samples (with or without cluster stars) are fully compatible. Also, 
once a suitable density model is found, the residual statistics of 
sample k125 (Figure 13b) do not show any significant feature that might signal 
undetected clumpiness left.


%

   \begin{figure}[htbp]
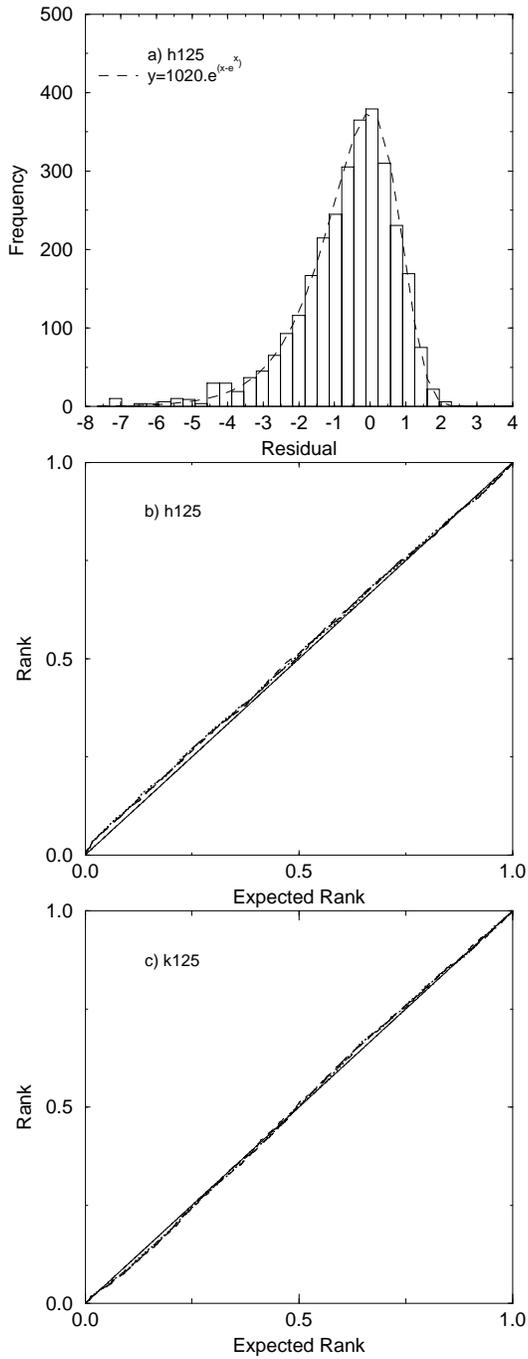
  
      \vspace{18.5cm}
	\special{ hscale=70 vscale=70 hoffset=0 voffset=-30.
	hsize=450 vsize=550 angle=0.  psfile=6440.f09 }
	\caption{Inverse  density residual  distribution for   the sample   h125.   
Residuals  are defined as y = ((single star volume) / average (101 stars)) 
{\bf a)} histogram of $log~y$.  The   smooth line is the  theoretical
distribution.  {\bf b)}  cumulative distribution  of residuals
(relative rank / expected rank of observed value). Departures from being parallel
to the diagonal would sign under or over frequency of residuals at relevant scales.
The presence of star clusters is visible  at bottom left. 
{\bf c)}  same as b), cluster stars  were excluded from the analysis (sample k125).  }  
	\label{MomoRes} 
   \end{figure}
%
%

This gives confidence that the bending is real and that figures 6cd, 7c
are the first "image" of  the galactic potential well.  The remarkable
feature here is that this well is extremely shallow.   We shall see in
the  following that any substantial layer  of dark matter contributing
significantly in the  local mass density  would  unmistakably sign its
presence  by deepening this well  far beyond observed shapes.  This is
true also for  samples H80 and H100 which, being dominated by  elder
stars are less affected by the clumpiness effect.


\section{Estimating $\rho_0$ }

Solving Eqs. (\ref{eq1}) and  (\ref{eq2}) for $\phi(z)$ can follow two
 different  ways:  a parametric  approach  and  a non  parametric one.
 Either were used on the above defined samples producing quite similar
 results.

\begin{itemize}

	\item In the parametric approach, realistic mathematical forms
are adopted for   both $\phi(z)$ and   $f(w_0)$.  Free parameters  for
$f(w_0)$ can be  determined by fitting  the observed vertical velocity
distribution then adopting $f(w_0)$ free parameters for the potential,
$\phi(z)$ can be derived from the  observed density.  The advantage of
this approach  is simplicity in so  far as manageable formulae  can be
found to represent  $f(w_0)$.  The cost is that  results may be biased
by the   choice of the model.   An  advantage may be   that systematic
effects can be traced and individual residuals can  be computed at the
position and  velocity of  each star.   This approach  is developed in
section 4.1.

	\item At  the  other  end  one can  try to  produce   a purely
numerical  fit.  Since this  kind of fit is  unstable to  noise in the
data, some  regularization should be   applied.  This  non  parametric
approach is developed in section 4.3.

\end{itemize}

\subsection {Parametric approach: velocity analysis }

The  observed W vertical velocity distributions  in the galactic plane
were derived  from Hipparcos proper motions   in galactic latitude and
parallaxes. For this analysis, samples  were restricted to stars below
$  b   =  \pm  10\degr$ so  that   $\mu_b/\pi$ simply   reflects the W
tangential velocity.

\begin{figure*}[htbp]
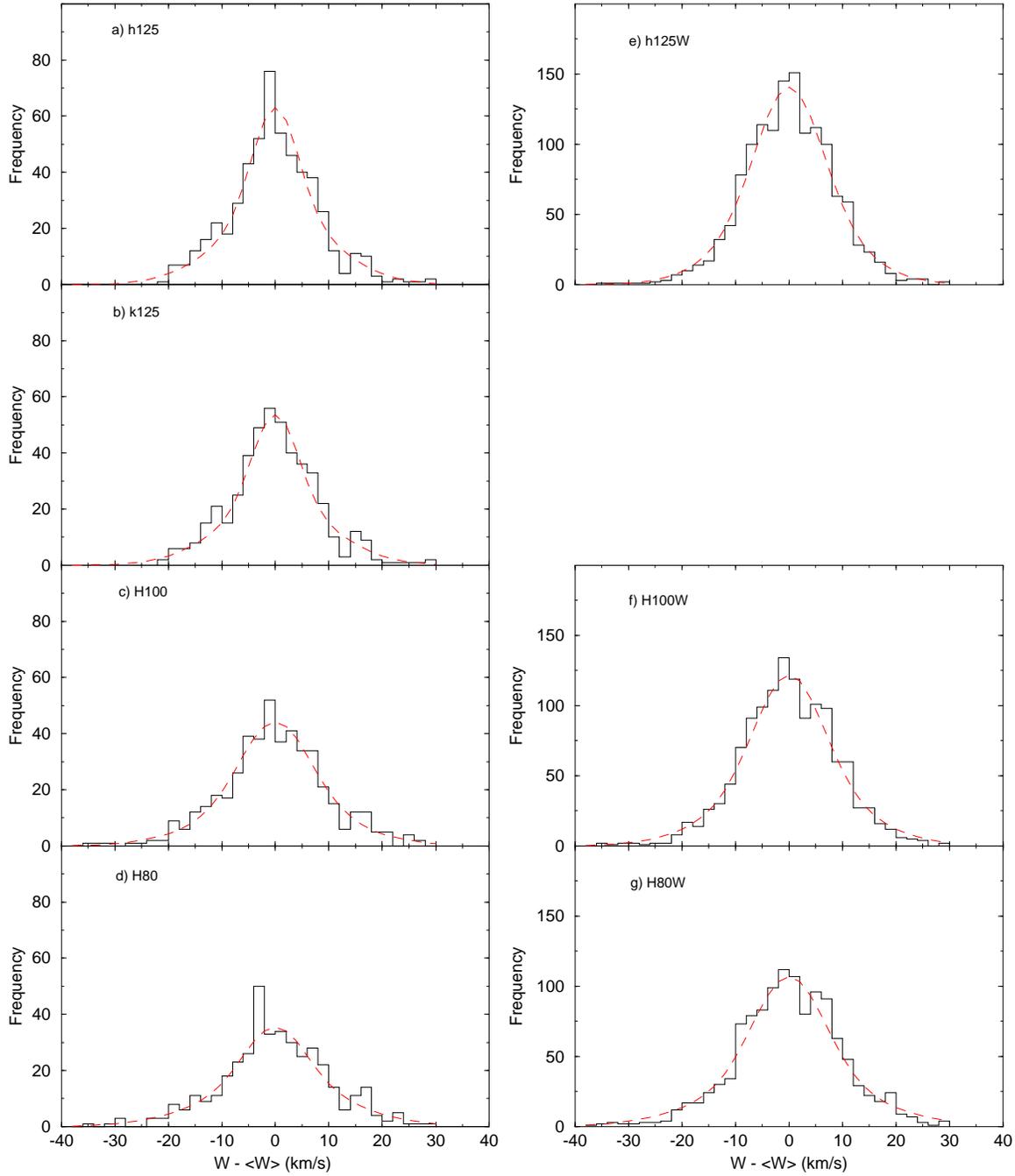

      \vspace{18cm}
	\special{ hscale=80 vscale=80 hoffset=-20 voffset=-40.
	hsize=450 vsize=550 angle=0.  psfile=6440.f10 }
      \caption{W  velocity  histograms  from $\mu_b$ of   low latitude
stars  (left hand column) and from  space velocities of all stars with
known radial  velocities  (right  hand column);   samples  : h125 {\bf
a)},{\bf e)} ; k125 {\bf b)} ; H100 {\bf c)}, {\bf f)} ; H80 
{\bf d)},{\bf g)}.}
         \label{Velhisto}
   \end{figure*}

W histograms   of the four  samples under  consideration are  shown on
Fig. \ref{Velhisto}.

%

%

%
%

All four distributions   show reasonable symmetry and smoothness  with
the following exceptions: h125 (Fig. \ref{Velhisto}a) shows pronounced
spikes corresponding to clusters   (spikes vanish in the histogram  of
k125) and  a dissymmetry of large velocities  (involving  only 5 stars
out of  456). The  plain  average $W_0$  has  been estimated   for all
samples and  appears  quite stable suggesting  that unsteady streaming
motions characteristics of  very young stars  do  not dominate.  Under
such  conditions, the capability of the  subsample of 536 low latitude
stars to  be representative of  the whole population  within 125 pc is
questionable.   Radial velocities have  been  compiled from the SIMBAD
database in order  to calculate space  velocities for as many stars as
possible. This resulted in a new velocity sample named h125W including
1366 stars spread in the whole sphere.  The distribution of W velocity
components has  been studied in the  following along the same lines as
the other velocity sample.

A double gaussian centered model was fitted to each centered histogram
using a simple maximum likelihood scheme:

the model fitted is

\begin{eqnarray}
	{\rm d} P(w) = \{{\beta \over \sqrt{2\pi}~\sigma_{w1}}
	\exp[-{1\over2}(w-<w>)^2/ \sigma_{w1}^2] \nonumber \\
	+ {1-\beta \over \sqrt{2\pi}~\sigma_{w2}} 
	\exp[-{1\over2}(w-<w>)^2/ \sigma_{w2}^2] ~\}~~{\rm d}w 
\label{eq18}
\end{eqnarray}

and the maximum likelihood is searched  for on $\beta$, $\sigma_{w1}$
and $\sigma_{w2}$.

There is no claim here that the kinematical mixture is isothermal. The
idea is simply that, if this mixture can  be accurately represented by
a  sum of gaussians,   then there is  a  simple  solution to  the self
consistent dynamical  equilibrium, the force law  and the local volume
density can  be solved for explicitly.  The justification lays in the
quality of the representation.  This is also justified a posteriori by
the fact that non parametric solutions do provide the same results.

 Table \ref{tab2} gives the  maximum likelihood solutions for the four
main samples  investigated here.  Two tests  of the quality of the fit
are  given, one  is the Kolmogorov-Smirnoff  test: Pk  is the observed
Kolmogorov maximum distance, Pk(0.8) and Pk(0.95) are the 20$\%$ and 5
$\%$ rejection thresholds for  this parameter. Obviously no  rejection
of the  double gaussian    representation  can be  supported  at   any
meaningful level.

                 
\begin{table*}
\begin{center}
\caption {Maximum Likelihood solutions for 2-gaussian fit of W velocity distribution.}
\label{tab2}
\small
\begin {tabular}{lrcccccccc}
\hline
\hline
 Sample & N & W$_{0}$&$\beta$ & $\sigma_{w1}$ & $\sigma_{w2}$& P$_{k}$&P$_{k}$(0.8)&P$_{k}$(0.95)\\
 &&km$\cdot s^{-1}$&&km$\cdot s^{-1}$&km$\cdot s^{-1}$&&&&\\
\hline
 h125 &536 & -6.53 &  0.60  &  10.40 &  4.46  &  .037 & .046 & .059\\
             &      &    (0.37) & (0.12) & (0.43) &(0.48) & & &&\\
\hline
h125W &1366 &   -6.94 &  0.50 &   10.67&   6.08 & .015 & .029 & .037\\
            &&    (0.23)  &(0.06)&  (0.31) & (0.27)&& &&\\
\hline
k125 &462  &   -6.83 &  0.60  &  10.4 &   4.20  & .019 & .050 & .063\\
              &&  (0.36) & (0.12) & (1.2) &  (1.5)&& &&\\
\hline
H100 &469  &   -6.39 &  0.70 &   11.46 &  4.79  & .016& .049 & .062\\
              &&   (0.46) & (0.15) & (0.46) & (0.63)&& &&\\
\hline
H100W &  1293  &  -6.93  & 0.50  &  12.49 &  6.39 &  .018 & .029 & .038\\
              &&  (0.27)  &(0.05) & (0.38) & (0.28)&& &&\\
\hline
H80  &407   &  -5.84 &  0.60  &  13.17 &  5.81  & .023 & .053 & .067\\
              &&  (0.49)  &(0.15) & (0.63)  &(0.68)&& &&\\
\hline
H80W  &      1214  &  -6.88  & 0.50  &  14.46  & 6.60  &  .024 &.031 & .039\\
               && (0.32) & (0.05) & (0.45) & (0.31)&& &&\\
\hline
\hline
\end{tabular}
\end{center}
\end{table*}

\begin{figure*}[htbp]
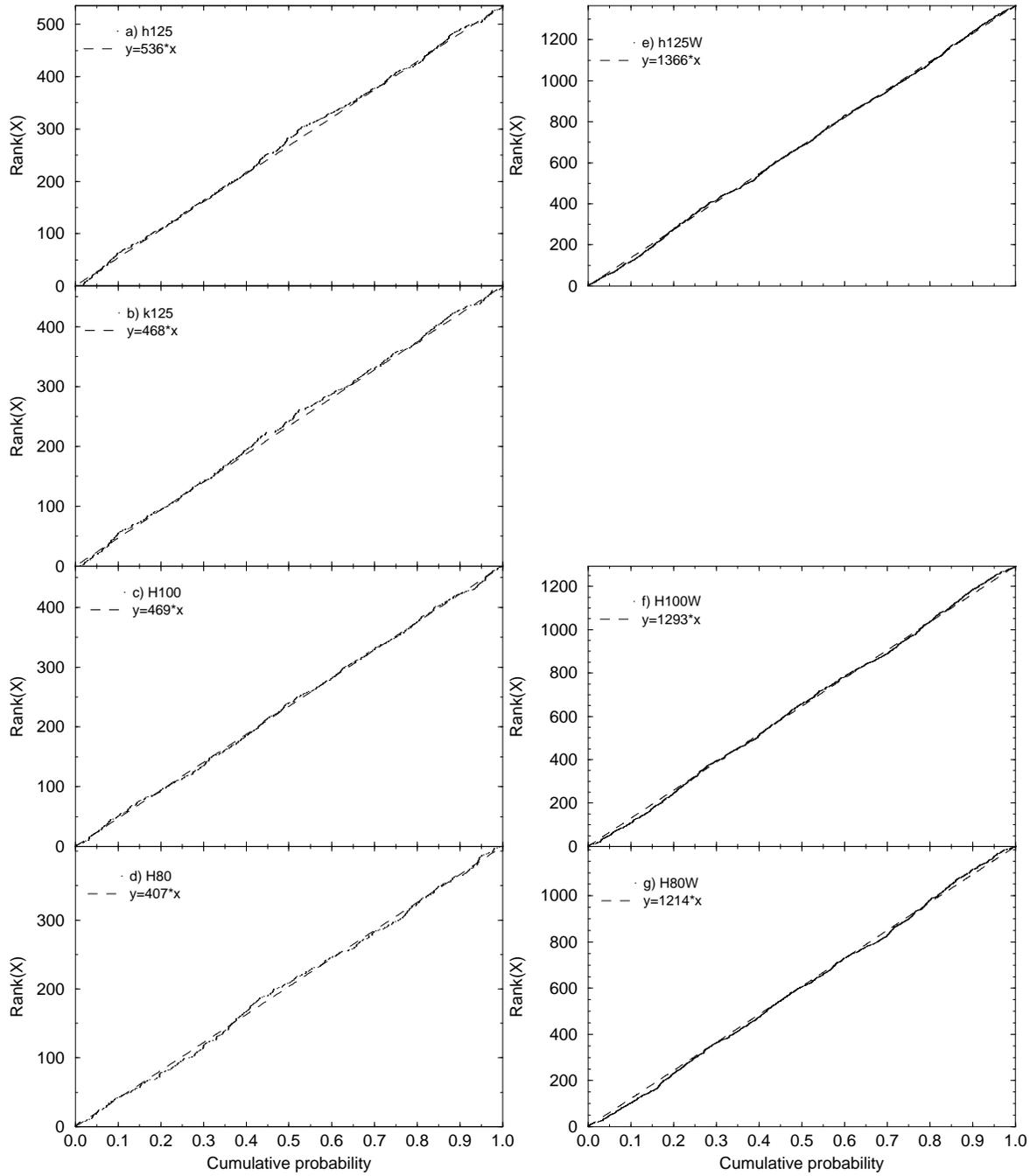

      \vspace{18cm}
	\special{ hscale=80 vscale=80 hoffset=-20 voffset=-40.
	hsize=450 vsize=550 angle=0.  psfile=6440.f11 }
      \caption{ W  velocity probability  plots (rank versus  estimated
cumulative probability) for samples of figure 10. Samples displaid just
as on figure 10.}
         \label{Velpplot}
\end{figure*}

The second test  is given  by probability  plots (Fig. \ref{Velpplot}), velocity
ranks in  each sample are  plotted  against the theoretical cumulative
probabilities   of associated   velocities: velocities  distributed
according  to  the estimated  distributions   would produce a strictly
linear  plot. So any  deviation from linearity signs  a failure of the
double          gaussian          to     represent       observations.

%

%
%
		
 The likelihood surface is regular around each maximum, the dependence
in  $\sigma_{w1}$ and   $\sigma_{w2}$  follows   a nice  parabola   in
log-likelihood indicating that  the  estimator behaves quite  alike  a
gaussian   variate  in this    region.   Standard errors follow.   The
proportion  $\beta$ is poorly  determined although different values of
$\beta$ may produce quite different  M.L.  estimates for $\sigma_{w1}$
and  $\sigma_{w2}$.  But the Likelihood  levels reached do not deviate
significantly from  oneanother. A series of  tests, not reported here,
show that adopting  any  of  those  solutions, nearly  equivalent   in
likelihood,  do produce extremely close  estimations  of the dynamical
density.  Expectedly,    what matters here,  is  the   quality of  the
distribution representation,  not the  choice of  specific  functions.
Eventually this   conclusion is comforted   by the excellent agreement
between the results of the parametric and non parametric approach.


\subsection {Parametric approach: density and potential analysis}

Once we  get models  fitting  W  velocity distributions, which  can  be
represented   by double gaussians,  the   self consistent solution  of
Eqs. (\ref{eq1}) and (\ref{eq2}) can be expressed explicitly as 


\begin{eqnarray}\
\nu(z) = \nu_0 ~ [\,\beta ~e^{- \phi(z) / \sigma_{w1}^2}
+ (1 - \beta)~ e^{- \phi(z) / \sigma_{w2}^2}\,]
\label{eq19}
\end{eqnarray}

Over the   very  small z  range  spanned by  our 125  pc sphere, the
potential well can be approximated at any required accuracy by a plain
quadratic form


\begin{equation}
\phi(z) = \alpha \cdot (z-z_0)^2 
\label{eq20}
\end{equation} 

So,  the density trend of  a tracer $\nu (z)$ can  be used to estimate
$\alpha$,   $\nu_0$  and   $z_0$,  once   $\beta$,  $\sigma_{w1}$  and
$\sigma_{w2}$  have been derived from  the velocities.  The likelihood
of a set of single star volumes under  the assumption that the density
trend   is  represented  by Eqs. (\ref{eq19}),    (\ref{eq20}), can be
expressed by Eq. (\ref{eq21}) similar to Eq. (\ref{eq5}).
 

\begin{equation}
\log L = \sum_{i=1,n}~(~\log \nu(z_i) - \nu(z_i)\cdot v_i~) 
\label{eq21}
\end{equation}

A maximum likelihood can be  searched for $\alpha$, $\nu_0$ and $z_0$.
According to the Poisson equation, the local dynamical density follows
from $\alpha$ as:
\begin{equation}
 \rho_0 = {1 \over 2 \pi G }~~~ \alpha 
\label{eq22}
\end{equation}  

Where   $1\,/(2\pi  G)    =  37.004$  if   $\alpha$  is   in   $ ({\rm
 km\,s}^{-1}\,{\rm    pc}^{-1})^2$   and $\rho_0$  in  $M_{\sun}\,{\rm
 pc}^{-3}$.

Table   \ref{tab3} summarizes the   results of solutions  based on the
three  main  samples  (the  separate  solutions derived   from samples
without identified cluster stars  do  not produce different  results).
Three  solutions are given  for each  sample:  one is the  3-parameter
M.L.    solution, the  other  two are   M.L.   solutions for $\alpha$,
$\nu_0$, while specific values have been forced for $z_0$.

%
%

\begin{table*}
\begin{center}
\caption {Maximum likelihood solutions for the local dynamical density from
the three Hipparcos samples.}
\label{tab3}
\small
\begin {tabular}{lrccccccc}
\hline
\hline
 Sample & N & z$_{0}$&$\beta$ & $\sigma_{w1}$ & $\sigma_{w2}$& 1/$\nu_{0}$ & $\rho_{0}$&Likelihood\\
 && pc && km$\cdot s^{-1}$ & km$\cdot s^{-1}$ &($\star$/pc$^{3})^{-1}$& M$_\odot$/pc$^{3}$ &\\
\hline
h125  &2977 & 12.0 & 0.50 & 10.67 & 6.08 & 1887.67 &  0.075 &  -25666.64\\ 
 \multicolumn{2}{c}{} &(6.0) &&&            &  (35.63)& (0.009)& \\
 \multicolumn{2}{c}{}&   0.0 & 0.50 & 10.67 & 6.08 & 1894.20 &  0.075 &  -25670.00\\
 \multicolumn{2}{c}{}& (forced) &&&            &  (35.70) & (0.010)& \\
 \multicolumn{2}{c}{}&  -11.0&  0.50 & 10.67 & 6.08 & 1926.14 &  0.059 &  -25676.31\\
  \multicolumn{2}{c}{}& (forced)  &&&            & (35.65) & (0.009)& \\
\hline
H100  &2677 &   4.0 & 0.50 & 12.49 & 6.39 & 1061.13 &  0.089 &  -21511.41\\ 
 \multicolumn{2}{c}{}&  (9.0)    &&&            &   (20.65)&  (0.017) &\\
 \multicolumn{2}{c}{}&   0.0 & 0.50 & 12.49 & 6.39 & 1061.52 &  0.089 &  -21511.71\\
 \multicolumn{2}{c}{}& (forced)  &&&            &  (20.65)& (0.017) &\\
 \multicolumn{2}{c}{}&-10.0&  0.50&  12.49 & 6.39 & 1079.97 &  0.064 &  -21513.69\\
 \multicolumn{2}{c}{}&(forced)   &&&            &   (20.87)& (0.016)&\\
\hline
H80   &2336 & $\>$13.0 & 0.50 & 14.46 & 6.60 &  631.57 &  0.076 &  -17485.54\\ 
 \multicolumn{2}{c}{}&   &&&                         &   (13.20)& (0.028) &\\
 \multicolumn{2}{c}{}&    0.0 & 0.50 & 14.46 & 6.60 &  642.95 &  0.046 &  -17487.20\\
 \multicolumn{2}{c}{}&(forced)  &&&            &   (13.38)&  (0.031) &\\
 \multicolumn{2}{c}{}&   -10.0 & 0.50 & 14.46&  6.60 &  654.98 &  0.040 & -17487.65\\
 \multicolumn{2}{c}{}&  (forced) &&&            &  (13.55)& (0.028)&\\
\hline
\hline
\end{tabular}
\end{center}
\end{table*}

All three parameter solutions produce  unexpected positive values  for
the z coordinate  of the galactic plane.  That  is,  the Sun would  be
very slightly south of the plane, in  contradiction with most accepted
values.  We do not consider  this result worth much consideration: for
one thing  the estimated local  density does not change appreciably if
$z_0$ is forced to 0 and even if it is pushed to a more classical $z_0
= -10$,  the estimated density decreases.  For  the other thing, there
is certainly  some clumpiness among the  youngest stars in this sample
which may blur  the shape of the  potential  well the more so  as this
well turns out shallower than  expected.  Since it  is not possible to
evaluate accurately  this effect we shall consider  it as an intrinsic
limitation of this determination.  Accordingly the adopted solution is
the weighted mean of  M.L.  solutions (weights proportional to inverse
square  standard errors),  but  error  bars are  extended  in order to
include all main solutions of Table \ref{tab3}.
                      
\begin{equation}
             \rho_0~=~0.076~\pm ~0.015~M_{\sun}~{\rm pc}^{-3} 
\label{eq23}          
\end{equation}

\begin{figure}[htbp]
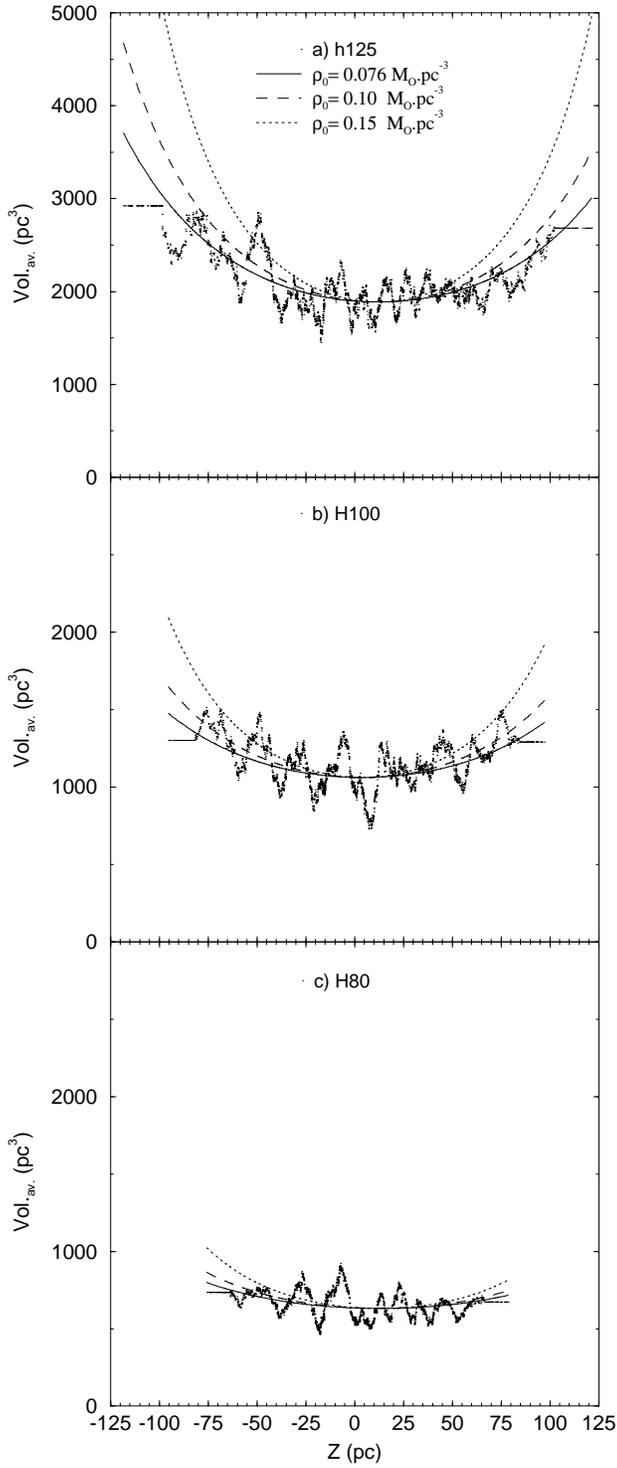

      \vspace{19cm}
	\special{ hscale=80 vscale=80 hoffset=-20 voffset=-50.
	hsize=450 vsize=560 angle=0.  psfile=6440.f12 }
      \caption{     Calculated inverse density    profiles under three
different   assumed     values  of    the    local   mass      density
$0.076~M_{\sun}~{\rm pc}^{-3}$ (best estimate),    $0.10~M_{\sun}~{\rm
pc}^{-3}$ and $0.15~M_{\sun}~{\rm   pc}^{-3}$ (from bottom  to top  on
each plot).  Based on Eq.  (\ref{eq20}) and the velocity distributions
of  sample  h125 {\bf  a)},  H100  {\bf  b)}  and  H80 {\bf  c)}.  For
comparison, the corresponding  moving  average  profiles are  repeated
from figures 6c, 7c, 8c.  }
         \label{Modfit}
   \end{figure}

Based on this  value  and on the  velocity distributions   obtained in
section 4.1, predicted density models defined by Eq. (\ref{eq19}) have
been overplotted on the z moving average  profiles of all three tracer
samples   (Fig. \ref{Modfit}     a,b,c).  For   comparison,   profiles
corresponding   to arbitrary  densities   $\rho_0~=~0.10~M_{\sun}~{\rm
pc}^{-3}$ and $\rho_0~=~0.15~M_{\sun}~{\rm pc}^{-3}$ are added. Obviously,  even
under quite   conservative  hypotheses, none of   the   samples can be
reconciled   with   such  local densities.    The  existence   of  any
substantial  disk shaped dark matter  would definitely  push all three
curves outside the acceptable range.


%

%
%

\begin{figure*}[htbp]
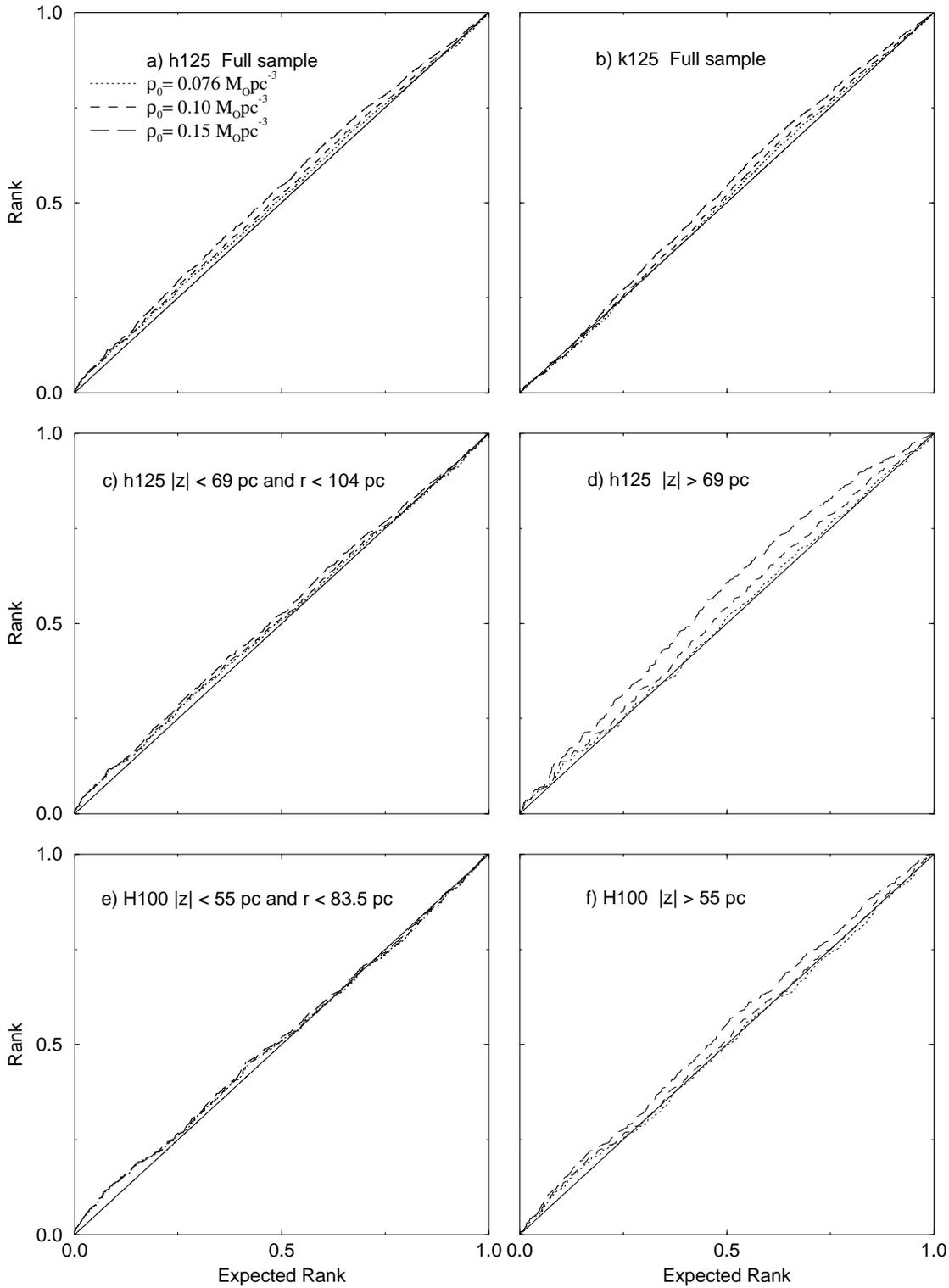

      \vspace{19cm}
	\special{ hscale=80 vscale=80 hoffset=0 voffset=-30.
	hsize=550 vsize=650 angle=0.  psfile=6440.f13 }
      \caption{Cumulative  distribution of residuals (relative rank versus
      expected relative rank at residual)  for $\rho_0=0.076$,  $\rho_0=0.10$ and
$\rho_0=0.15$.  Departures to the diagonal  sign anomalies in residual
frequency. Full samples  h125 {\bf a)} and k125 {\bf b)}.
 Sub-samples h125 {\bf c)} $ \vert z \vert \leq 69~{\rm pc}$ and
$r<104~{\rm pc}$ ; {\bf d)} $\vert z\vert >69~{\rm pc}$.
Sub-samples H100 {\bf e)} $\vert z \vert \leq 55~{\rm pc}$ and 
$r< 83.5~{\rm pc}$; {\bf f)} $\vert z \vert > 55~{\rm pc}$. }
         \label{Modres}
   \end{figure*}

The distributions  of  single star volume residuals  under  the three
density   models have  been   scrutinized  against  possible unmodeled
systematic effects.  Probability plots  as defined in section 3.3 (Eq.
(\ref{eq11})) are  given in  Fig. \ref{Modres}.  
 
The first two plots on top show the probability plot of residuals 
for samples h125 and k125 . Clearly enough models based on local mass
densities  higher than 0.08 generate a systematic distortion of the 
cumulative distribution of residuals with respect to the expected one. 
There is a small excess of low residuals even with our best fit model,
but this effect vanishes when cluster stars are removed (sample k125)
This correction does not change the conclusions on the local mass. 
 
At this stage , the validity of the simplified assumption made on the
potential can also be tested. A square function in z has been adopted for the
potential, meaning a linear $K_z$. This means some homogeneity of the mass
distribution over a 100 parsecs scale. If a substantial fraction of the mass 
happened to be concentrated in a very thin layer, this approximation would 
not hold.
 The Hipparcos samples are the first able to test this hypothesis
over such small scales. The test is provided by the statistics of single
star residuals. Would the $K_z$ slope change significantly within the limits 
of the sample, then this statistics of subsamples at low z would not
match the expectation for the same assumed $\rho_0$. This test is
presented in the bottom parts of Fig.    \ref{Modres}. 
Samples h125 and H100 were split in different  subsamples selected by 
$z$ after restriction to a cylinder  of radius  $r$. Expectedly, the failure
 of models based on $\rho_0 = 0.10$ and beyond is enhanced at high z
($|z|>  69~{\rm pc}$). 
But,  even over the restricted range $|z|<  69~{\rm pc}$ there  is signal left in
the residuals to reject high density solutions. The low z part
of sample H100 ($|z| <  55~{\rm pc}$) is unable to produce any discriminant signal.
 
These residual plots fully confirm the  result that local total volume
densities  beyond $0.10~M_{\sun}~{\rm  pc}^{-3}$  cannot be reconciled
with  the  observed  velocity-density  distributions.   Similar  plots
restricted to subsamples selected by  distance or magnitude or  colour
do  not show significant  deviations that might  sign  the presence of
major selection effects in the sampling.
 
In terms of mass model the assumed linear $K_z$ implies that the total
mass density is constant close to the plane.  Assuming that it is true
only for half of the  mass, for instance  the stellar distribution and
that the gas distribution follows a sharper gaussian distribution with
a scale height of 75 pc we may determine the extra forth order term in
the potential. Determining the amplitude of the potential on counts at
75 pc leads to   a  very small  difference   of 1/24 between  the  two
models. The resulting error on the derived $\rho_0$ is negligible with
respect to other uncertainties.

%
%
%
\subsection {Non parametric approach}  
The  non  parametric approach   does   not rely upon   any  particular
mathematical model  for  the  velocity distribution  function  nor the
height density profile (\cite{Pichon97}).   That is: ideally we should
try and  solve self consistently Eqs.  (1)  and (2) for $\rho_0$, with
functions $ f(|w_0|)  $ and $\nu(\phi)$ defined  on a purely numerical
basis  (in the range investigated   here, approximating $\phi(z)$ by a
simple square  function is well justified).   However a pure numerical
solution is unstable: a small  departure in the measured data  (e.g.\
due to noise) may produce  drastically different solutions since these
solutions are dominated   by artifacts due   to  the amplification  of
noise.  Some  kind of trade  off must  therefore  be found between the
level of  smoothness  imposed on the solution  in  order to deal  with
these  artifacts  on the  one  hand,  and  the level   of fluctuations
consistent with the amount  of information in  the signal on the other
hand.   Finding  such  a  balance   is  called the  ``regularization''
(\cite{WW79}) of the self  consistency problem.  Regularization allows
to avoid over-interpretation of the data.  Here we chose to regularize
the non-parametric  self   consistency problem by  requiring  a smooth
solution.  This criterion  is tunable  so that  its  importance may be
adjusted  according  to  the noise  level  (i.e.\  the  worse the data
quality is, or  the sparser the  sample is, the lower the  informative
contents of  the  solution will  be  and  the  smoother  the  restored
distribution will be).

Under  the assumption that  the potential is separable, the proposed
method yields a unique mean density compatible with the measurement of
the velocity distribution perpendicular to  the Galactic plane and the
corresponding measured density profile.

\paragraph{The discretized integral equation}

From the  knowledge of the  level of  errors  in the  measurements and
given the  finite  number of  data points,  it  is possible  to obtain
$f(w)$  by  fitting  the data   with  some model.  A  general approach
assumes that the solution  can be described  by its projection onto  a
complete      basis    of  functions with finite 
support (here cubic splines)    $\{e_k(w);    k=1,\ldots,N\}$:
\begin{equation}
	f(w) = \sum_{k=1}^N f_k \; e_k(w) \equiv {\mathbf f} \cdot
	{\mathbf e}  \;. 
	\label{e:f-model-1}
\end{equation}
where the $N$ parameters to fit are  the weights $f_k$. 
Real data correspond to discrete measurements $z_i$ and ${w}_j$ of $z$
and  $w$   respectively.   Following   the   non-parametric  expansion
in Eq.~(\ref{e:f-model-1}), Eq.~(1) becomes:
\begin{equation}
	\rho_{i} \equiv \rho(z_i) =  \sum_{k=1}^N  A_{i,k} (\alpha,z_0) \;
		f_k \, , \label{e:Fphi-model} 
\end{equation}
with
\begin{equation}
 	A_{i,k}(\alpha,z_0) = \int_{\sqrt{2 \alpha z^2_{i} }}^{\infty}
	\frac{e_k(w) w} 
	{\sqrt{w^{2}- 2 \alpha z^2_{i}}}\,\d{w}\,.
	\label{e:Fphi-model-coef} 
\end{equation}
Since  the  relation  between   $\M{\tilde \rho }\equiv \{ \rho_{i} \} $
and   $
\M{f} \equiv \{f_k\}$ 
 is linear, Eq.~(\ref{e:Fphi-model}) ---  the discretized
form of the integral equation~(1)  ---  can  be  written  in a
matrix form:


\begin{equation} 
   \M{\tilde \rho} = \M{A}(\alpha,z_0) \mdot \M{f} 
	\label{e:Fphi-model-2}
\end{equation}
Here $z_0$ accounts for the relative  offset in the position of the sun
with respect to the galactic plane as defined by the Hipparcos sample to
the measured density.

The Hipparcos sample gives access within a given sphere to a number of
stars for which the velocity, $w$, and/or the altitude, $z$, is known.
In other words the measurement yields a  realization of the cumulative
height density distribution  and the cumulative velocity distribution.
Indeed for  each star at  height $z$, within $\d  z$ of the  next star
(once they  have been sorted  in  height), we  can associate a  volume
given by


\begin{equation}
\pi \, \d z \, R^2 (1-z^2/R^2)
        \label{e:vcor}
\end{equation}
 where  $R$ is  the radius of    the Hipparcos sphere;  an alternative
estimated local density for this star is  given by the inverse of this
volume.  This  density estimator yields  a local density consistent
with that derived  in section 3.   An estimate of  its cumulative
distribution  is directly  available   from the   measurements without
resorting to  averaging nor binning.  In order  to project the modeled
distribution, $f$, into data space it is  therefore best to model both
the cumulative    density  distribution, $R(z)$   and  the  cumulative
velocity distribution, $F(w)$, which satisfy respectively


\begin{equation}
	R(z)= \int_{-\infty}^z \rho(z) \, \d z \,,
\quad {\rm and}  \quad 
	F(w)= \int_{-\infty}^w f(w) \, \d w \,.
	\label{e:cumul}
\end{equation}

Now if $f(w)$ obeys \Eq{f-model-1}, its cumulative distribution will satisfy


\begin{equation}
	F(w) = \sum_{k=1}^N f_k \; E_k(w)  \equiv {\mathbf f} \cdot
	{\mathbf E}  \;, 
	\label{e:F-model-1}
\end{equation}
where the basis of  functions $\{E_k(w); k=1,\ldots,N\}$ is simply the
anti-derivative of $\{e_k(w)\}$ with respect to $w$.

Similarly the cumulative density distribution satisfies


\begin{eqnarray}
	R_{i} \equiv R(z_i) =  \sum_{k=1}^N \frac{1}{\sqrt{2\alpha}}
C_{k} \left[
\sqrt{2\alpha}(z_i-z_0)\right] \; f_k \, , \nonumber \\
\quad
{\rm with} 
\quad
 	C_{k}(Z) =
\int_0^Z \int_{z}^{\infty}
	\frac{e_k(w) w} 
	{\sqrt{w^{2}-  z^2}}\,\d{w} \, \d{z}\,.
	\label{e:Fphi-model-coef-2} 
\end{eqnarray}
Note that for a cubic spline basis,  the function $C_k$ is algebraic (though
somewhat clumsy).  \Eq{F-model-1}   and \Eq{Fphi-model-coef-2} provide 
two  non  parametric means  of relating  the  measurements,  $R_i$  and $F_i$  
to  the underlying distribution, $f(w)$, with two extra parametres, $z_0$ and $\alpha$,
which one should adjust so that the distributions match.

\paragraph{Practical implementation}

First we compute the  best regularized fit  of the cumulative velocity
distribution  (we  use  $\chi^2$  fitting even   though  the noise  is
Poissonian   because  it  yields   a simpler   quadratic  problem). We
therefore minimize
 the quadratic form:


\begin{equation}
	Q_\R{\mu}(\M{f}) =
		\T{(\tilde{\M{F}} -\M{B}\mdot \M{f})} \mdot \M{W}
		\mdot (\tilde{\M{F}} -  \M{B}\mdot  \M{f})
		+ \mu \, \T{\M{f}} \mdot \M{K} \mdot \M{f} \, ,
	\label{e:Q-quad}
\end{equation}
\noindent where  $\tilde{\M{F}}$ is  the measured cumulative  velocity
distribution, $\M{K}$ the  discrete 2nd order differentiation
operator,
and $\M{B}=\{E_k(w_i)\} $  ;  the
weight matrix $\M{W}$  is the inverse  of the covariance matrix of the
error in the 
data and  $\mu$ is the  Lagrange parameter  
tuning the  level of regularization.
   Error estimation is achieved via  Monte  Carlo simulation: the
regularization parameter is initialized  at   some fixed value   (here
$\mu=10$);  from  this  value   a first  cumulative   distribution  is
fitted.  A draught  corresponding   to  the same   number of  measured
velocity  is generated and the  procedure iterated. An estimate of the
error bars  per point  follows.  The  best fit  $\M{f}_\R{\mu}$  which
minimizes $Q_\R{\mu}$ in Eq.~\ref{e:Q-quad} is:


\begin{equation}
	\M{f}_\R{\mu} = (\T{\M{B}}\mdot \M{ W} \mdot \M{B}   + \mu\,\M{K})^{-1}
\mdot		\T{\M{ B}}\mdot \M{W} \mdot \tilde{\M{F}}\,.
		\label{e:Q-quad-solution} 
\end{equation}
Next we 
apply  \Eq{Fphi-model-coef-2} 
to project  the distribution function
into cumulative density space. Calling  $\M{C}$ the matrix given
by its components $C_{i,k}= C_k \left[
\sqrt{2\alpha}(z_i-z_0)\right]/\sqrt{2 \alpha}$,
 we 
 then adjust  the  potential curvature $\alpha$
(together with the possible offset $z_0$) while minimizing


\begin{eqnarray}
	| \M{\tilde R} -  \M{C}(\alpha,z_0) \cdot  \M{f}_\mu |^2 = \hspace{4cm} 
\nonumber \\ 
	| \M{\tilde R} -  \M{C}(\alpha,z_0) \cdot
 	(\T{\M{B}}\mdot \M{ W} \mdot \M{B}   + \mu\,\M{K})^{-1}
	\mdot		\T{\M{ B}}\mdot \M{W} \mdot \tilde{\M{F}}\, |^2
	\, , \label{e:minim}
\end{eqnarray}
with respect  to  $\alpha$ and    $z_0$.
   This 
 yields the  local density of the
disk.  Note that the last two steps could be carried simultaneously,
i.e. we could seek the triplet, $(\mu,\alpha,z_0)$ 
which minimizes \Eq{minim}, or even better, 
the unknown, $ \mu, \alpha,z_0$ and $\{f_k\}, \; k=1\cdots N$, 
which minimize the quantity: 


\begin{eqnarray}
		\T{(\tilde{\M{F}} -\M{B}\mdot \M{f})} \mdot \M{W}_1
		\mdot (\tilde{\M{F}} -  \M{B}\mdot  \M{f})
	+ \hspace{2cm} \nonumber \\
		\T{(\tilde{\M{R}} -\M{C}\mdot \M{f})} \mdot \M{W}_2
		\mdot (\tilde{\M{R}} -  \M{C}\mdot  \M{f})
		+ \mu \, \T{\M{f}} \mdot \M{K} \mdot \M{f} \, ,
	\label{e:Q-quad-2}
\end{eqnarray}

though
in practice the  level of regularization,  $\mu$, -- within some range,  only
 affects the wings  of the  reconstructed  density profile and  this has  no
 consequence on the central density value, $\rho_0$ (a 
supplementary difficulty
 involved in minimizing \Eq{Q-quad-2}
would be to assess the relative weights between 
errors in position and velocity).

Since our purpose  was primarily to validate the parametric approach 
described in sections 4.1 and 4.2, we also minimized
\begin{equation}
| \M{\tilde \rho} -  \M{A}(\alpha,z_0) \cdot  \M{f}_\mu |^2
\end{equation}
instead of \Eq{minim} using the nearest neighbour density estimator
described in section 3.
This lead to  non parametric estimator for
sample h125 given by
$\rho_0= 0.07-0.08~ M_{\sun}~{\rm pc}^{-3} $ with $z_0= 13 \, \, {\rm pc}$.
The analysis of sample H100 gives~
$\rho_0= 0.08-0.10~ M_{\sun}~{\rm pc}^{-3} $.
In  practice, the parametric   and  non parametric  analysis 
therefore  lead to
statistically consistent answer.

\subsection {Radial force correction}
   The $K_z$ (vertical forces) determination constrains the local mass
  density:  precisely $\partial^2 \Phi_{total} \over \partial z^2$.
   A complete determination of the Poisson
  equation should include the $K_r$ (radial force) component.  This
  second contribution is small and can be estimated from the A and B
  Oort's constants:
\begin{equation}
  \Delta \Phi_{total} = {\partial^2 \Phi_{total} \over \partial
  z^2} + 2(B^2-A^2) = 4\pi G \rho_{total}
\label{eq35}
\end{equation}
  From the dynamical analysis of the present Hipparcos sample, we
  have obtained 0.076 for the $\partial^2 \Phi_{total}\over\partial
  z^2$ term.  The term including A and B Oort's constants ranges
  between $\pm0.004$; it is probably positive corresponding to a
  rotation curve locally decreasing. This term is small and
  we assume it is zero to simplify the discussion.

\section{Conclusions}  

The consequence of estimations given   in previous sections is to  fix
the total mass  density in the  galactic disc in the  neighbourhood of
the Sun in the range 0.065-0.10~$M_{\odot}~{\rm pc}^{-3}$ with a most
probable value  by 0.076~$M_{\odot}~{\rm  pc}^{-3}$.  In conclusion we
review below the consequences of this very low value in terms of local
dynamical  mass versus  observed density,   in  terms of  global
galaxy modeling and in terms of dark matter distribution.

\subsection{Dynamical mass versus known matter}

Previous dynamical determinations of the local mass density used to be
based  on star counts  a few  hundred parsecs  away from  the galactic
plane providing constraints  for  the local column  density integrated
over a substantial layer.   Comparisons were tentatively made with the
observed density of  known galactic components, namely visible stellar
populations,    stellar remnants, ISM.   So,   on  either side of  the
comparison  there were   assumptions on the   vertical distribution of
matter involved.

   For the first time, the  present determination is based on strictly
   local data  inside  a sphere of more   than  100~parsecs radius and
   gives access directly  to the total mass density  around the Sun in
   the galactic plane.

   On  the side  of the known  matter, the current  situation is the
   following: the main  contributors are  main  sequence stars of  the
   disc, stellar remnants and the interstellar medium .

   1)  Recent determinations  of the stellar  disc luminosity function
   have   a maximum near $M_v$=12  and  decreases quickly beyond. 
   Counting nearby stars Wielen et al. (1983) estimate the total mass
density of main sequence stars in the solar neighbourhood as
0.037 plus 0.008 in the form of giants and white dwarfs. Since then the
luminosity function below $M_v$ = 13 has been revised downwards (Reid et
al 1995, Gould et al 1996). Thus main sequence stars are expected to
contribute by 0.027-0.033$~M_{\odot}~{\rm pc^{-3}}$.
This  is the most  reliable part of the  known  matter, yet some extra
uncertainty might result from the non detection of low mass binaries.
Preliminary results based on Hipparcos parallaxes seem to imply
that stellar distances are on the average slightly larger than expected
which would mean an even smaller density (Jahreiss and Wielen, 1997).

   2) Stellar remnants dominated by  white dwarfs should contribute by
   an   additional $0.015  \pm 0.005~M_{\odot}~{\rm pc^{-3}}$. This
   figure  is compatible with  observational determination of the WD's
   luminosity function (\cite{Sion77}, \cite{Ishi82}).  It
   does not   conflict with  state   of  the art scenarios    of  star
   formation.

   3)  The   local mass density   of  the ISM  estimated from   HI and
   $\rm{H_2}$    abundances.        $\rm{H_2}$     is    deduced  from
   $\rm{CO}$/$\rm{H_2}$  ratio.   This  estimation  suffers very large
   uncertainties.   According to   \cite{Com91}, a contribution  about
   0.04 is  reasonable, but  errors by a  factor two  or more are not
   excluded.  This is certainly the weakest link of this analysis.

   So, 0.085   $M_{\odot}~{\rm  pc^{-3}}$ is  probably   an acceptable
   figure for  the known mass.   But the  uncertainty  is larger  than
   0.02.    It does not make sense   to overcomment about the observed
   mass  density being larger  than   the dynamical estimate: for  one
   thing the  discrepancy is far below the  error  bars, for the other
   thing   there has been little  effort  dedicated so far to estimate
   properly the smallest  density  compatible with  observations,  the
   attention being always focused on upper bounds.  For the moment the
   only conclusion is  that the density   of known matter has  a lower
   limit of say  0.065~$ M_{\odot}~{\rm pc^{-3}}$.   In term  of mass
   discrepancy   controversy, the  main  uncertainty   is now on   the
   observation side.

  \subsection{Galactic mass model}

   In the  following we adopt  a simple, two component modelisation of
  the   galactic  disc  including   stars  and     interstellar medium
  (double-exponential    and    exponential-sech-square   laws).   The
  formalism  is  standard,  it is   for instance  the  one adopted  by
  Gould et al (1996) (Eq.  1.1 in their paper).

  A good convergence  is now obtained on   the main parameters  of the
disc  populations although   there is no  agreement  as  to  the exact
decomposition in populations (see for example, \cite{Reid83}, Robin et
al 1986ab-1996, Haywood 1997ab, Gould et al 1996, etc..).

  Based on   the  discussions quoted   above  the tentative structural
parameters of our model are close  to Gould's (see also \cite{Sac97}):
stellar disc scale  length 2.5\,kpc, scale  height 323\,pc; thick disc
scale length 3.5\,kpc, scale height  (exponential) 656\,pc.  20$\%$ of
the local stellar  density is  in  the thick disc.   The  error on the
scale length    has  little   effect   on    the disc   mass    $M_d$:
$M_d(2.5{\rm\,kpc})/M_d(3.5{\rm\,kpc})=1.34$ (assuming $R_0$=8.5\,kpc)
and  a factor 1.26  on the   amplitude  of the  corresponding velocity
curves.

  Adopting a total stellar mass density $\rho_\star=.043M_\odot\, {\rm
pc}^{-3}$,       the   surface     density    at     $R_0$    is  then
$\Sigma_0=33.4~M_\odot\,{\rm pc}^{-2}$ 
  
  The   contributions of the bulge and   stellar halo are small beyond
3~kpc of the galactic center. They are neglected in this discussion.
   
  The interstellar  matter is accounted  for  by a  double exponential
  disc 2500/80~pc  its  local  density  is set  to  0.04~$M_\odot~{\rm
  pc}^{-3}$.

  The   total column  density due  to    gas and stellar  component is
  $\Sigma_0=40~M_{\odot}~{\rm pc}^{-2}$.

  A dark halo is necessary to maintain a flat rotation curve beyond
  the solar galactic radius.  It is modeled with a Miyamoto spheroid
  that allows flattening.

  This mass model  is fitted to the rotation  curve  setting the solar
  galactic radius to $R_0$=8.5~kpc and the velocity  curve at $R_0$ to
  $220\,{\rm km\,s^{-1}}$.  The velocity  curve is assumed flat beyond
  5~kpc of the center to 20~kpc.

  The first consequence  is that the stellar disc (stars+remnants+gas)
  does not contribute   for more than half   the mass implied by   the
  rotation   curve at $R_0$=8.5~kpc:  the  galactic  disc  is far from
  maximal.   Even  with $R_0$=7.5\,kpc, a maximal   disc  would mean a
  rotation curve plateau as low  as 164~km\,s$^{-1}$. This is excluded
  by inner HI and CO rotation curves.

  Filling  the galactic mass  distribution  with a strictly  spherical
  dark  matter halo required  by  the rotation curve,  the dark matter
  local                  density           comes      out           as
  $\rho_{dark\,halo}(R_0,z=0)=0.007~M_{\odot}~{\rm pc}^{-3}$.   Such a
  local  density  of dark  matter is  well  compatible  with the above
  quoted uncertainties of the  dynamical and known mass.  However, any
  attempt to  flatten  the  mass distribution   of the  dark component
  result  in an increase of  the local halo  density.   Under the most
  extreme hypotheses,  that   is considering the   interstellar matter
  contribution negligible and giving   the stellar component a  total
  density $0.043~M_{\odot}~{\rm  pc}^{-3}$ with  the same scale length
  and  adopting our best dynamical  estimate at its face value (0.076)
  the acceptable halo scale height cannot go below 2150~pc.

 So, there  is   just  room for  a   spherical  halo   (local density:
$0.008~M_{\odot}~{\rm pc}^{-3}$).  Extreme  changes of  the parameters
are needed to permit a significantly flatter dynamical halo.  There is
no room  for an important  amount of  dark matter in  the disc.   Dark
matter models  assuming  that the dark matter   is in form of  a  flat
component    related to  a  flat fractal    ISM  should be modified to
equivalent models where  dark matter is in  form of fractal structures
distributed in  the  halo (Pfenniger  et  al  1994ab, Gerhard \&  Silk
1996).
  
\subsection{Summary of conclusions}
  \begin{enumerate}
  
  \item The potential well across the galactic plane has been traced
  practically hypothesis-free and model-free, it turns out to be shallower 
  than expected.

  \item   The   local    dynamical  volume  density     comes   out as
$0.076\pm0.015~M_{\odot}~{\rm pc}^{-3}$ a  value well compatible  with
all existing observations of the known matter.

  \item  Building a  disc-thick disc mass   model compatible with this
constraint  as well as  previous  determinations of  the local surface
density, it is shown that  such a  disc  cannot be maximal, a  massive
halo of dark matter is required.
  
  \item The dark halo should be spherical or nearly spherical in order
for its local density to remain within the range permitted between the
known matter density and   the current determination of the  dynamical
density.
  
  \item There is no room left for any disk shaped component of dark matter.
  \end{enumerate}
   
\begin{acknowledgements}
Les travaux de  MC \`a l'IUP  de  Vannes ont \'et\'e rendus  possibles
gr\^ace  \`a  une  aide sp\'ecifique de   la  Mission Scientifique  et
Technique (DSPT3) et \`a l'assistance permanente de Robert Nadot.
CP thanks E.~Thi\'ebaut and O.   Gerhard for valuable discussions  and
acknowledges funding from the Swiss NF.
\end{acknowledgements}

\end{document}